\documentclass[12pt]{article}
\usepackage{times}
\usepackage{geometry}
\usepackage[utf8]{inputenc}
\usepackage{enumitem,amssymb}
\usepackage{ragged2e}
\usepackage{graphicx}
\usepackage{xcolor}
\usepackage{fancyhdr}
\usepackage{verbatim}
\usepackage[
    natbib=true,
    style=numeric,
    sorting=none
]{biblatex}
\newlist{thematic}{itemize}{8}
\setlist[thematic]{label=$\square$}
\usepackage{pifont}

\begin{document}
\raggedright
\huge{
Investing in the Unrivaled Potential of Wide-Separation Sub-Jupiter Exoplanet Detection and Characterisation with JWST}
\large \\
\vspace{4mm}
Aarynn L. Carter$^1$, Rachel Bowens-Rubin$^2$, Per Calissendorff$^3$, Jens Kammerer$^4$, Yiting Li$^3$, Michael R. Meyer$^3$, Mark Booth$^5$, Samuel M. Factor$^6$, Kyle Franson$^6$, Eric Gaidos$^7$, Jarron M. Leisenring$^8$, Ben W.P. Lew$^9$, Raquel A. Martinez$^{10}$, Isabel Rebollido$^{11}$, Emily Rickman$^{12}$, Ben J. Sutlieff$^{13}$, Kimberly Ward-Duong$^{14}$, Zhoujian Zhang$^2$\\

\vspace{4mm}
{\small$^1$ Space Telescope Science Institute, 3700 San Martin Dr, Baltimore, MD 21218, USA \\
$^2$ Department of Astronomy \& Astrophysics, University of California, Santa Cruz, Santa Cruz, CA 95060 \\
$^3$ Department of Astronomy, University of Michigan, Ann Arbor, MI 48109 \\
$^4$ European Southern Observatory, Karl-Schwarzschild-Str. 2, 85748 Garching, Germany \\
$^5$ UK Astronomy Technology Centre, Royal Observatory Edinburgh, Blackford Hill, Edinburgh EH9 3HJ, UK \\
$^6$ Department of Astronomy, The University of Texas at Austin, Austin, TX 78712 \\
$^7$ Department of Earth Sciences, University of Hawai’i at Manoa, 1680 East-West Rd, Honolulu, HI 96822, USA \\
$^8$ Department of Astronomy / Steward Observatory, University of Arizona, Tucson, AX 85721 \\
$^9$ Bay Area Environmental Research Institute / NASA AMes Research Center, P.O. Box 25 Moffett Field, CA, 94035 USA \\
$^{10}$ Department of Physics \& Astronomy, University of California, Irvine, Irvine, CA 92697 \\
$^{11}$ European Space Astronomy Centre (ESAC),European Space Agency (ESA), Camino Bajo del Castillo s/n, Villanueva de la Cañada, 28692, Madrid, Spain \\
$^{12}$ European Space Agency (ESA), ESA Office, Space Telescope Science Institute, 3700 San Martin Drive, Baltimore, MD 21218, USA \\
$^{13}$ Institute for Astronomy, University of Edinburgh, Royal Observatory, Blackford Hill, Edinburgh, EH9 3HJ, United Kingdom \\
$^{14}$ Department of Astronomy, Smith College, Northampton, MA, 01063 \\
}
\vspace{4mm}
\noindent \textbf{Thematic Areas (Check all that apply):} \linebreak $\boxtimes$ (Theme A) Key science themes that should be prioritized for future JWST and HST observations 
\linebreak $\square$ (Theme B) Advice on optimal timing for substantive follow-up observations and mechanisms for enabling exoplanet science with HST and/or JWST \linebreak
$\square$ (Theme C) The appropriate scale of resources likely required to support exoplanet science with HST and/or JWST 
 \linebreak
$\boxtimes$ (Theme D) A specific concept for a large-scale ($\sim$500 hours) Director’s Discretionary exoplanet program to start implementation by JWST Cycle 3.
 \linebreak

\clearpage
\justify{
\textbf{Summary:} 
We advocate for a large scale imaging survey of nearby young moving groups and star-forming regions to directly detect exoplanets over an unexplored range of masses, ages and orbits. Discovered objects will be identified early enough in JWST's lifetime to leverage its unparalleled capabilities for long-term atmospheric characterisation, and will uniquely complement the known population of exoplanets and brown dwarfs. Furthermore, this survey will constrain the occurrence of the novel wide sub-Jovian exoplanet population, informing multiple theories of planetary formation and evolution. Observations with NIRCam F200W+F444W dual-band coronagraphy will readily provide sub-Jupiter mass sensitivities beyond $\sim$0.4$''$ (F444W) and can also be used to rule out some contaminating background sources (F200W). At this large scale, targets can be sequenced by spectral type to enable robust self-referencing for PSF subtraction. This eliminates the need for dedicated reference observations required by GO programs and dramatically increases the overall science observing efficiency. With an exposure of $\sim$30 minutes per target, the sub-Jupiter regime can be explored across 250 targets for $\sim$400 hours of exposure time including overheads. An additional, pre-allocated, $\sim$100 hours of observing time would enable rapid multi-epoch vetting of the lowest mass detections (which are undetectable in F200W). The total time required for a survey such as this is not fixed, and could be scaled in conjunction with the minimum number of detected exoplanet companions.}


\pagebreak
\justify{

\textbf{Anticipated Science Objectives:}
JWST direct imaging presents the \textit{only} opportunity in the foreseeable future to detect \textit{and} characterise the sub-Jupiter exoplanet population beyond 10~au (Fig.\,1). By surveying $\sim$250 targets across several young stellar groups, we expect a yield of at least 10, and as many as 70, sub-Jupiters (assuming planet mass distributions based on \cite{Fult21}). This sample is large and diverse enough to enable statistical occurrence rate studies, and will facilitate detailed follow-up investigations on discovered objects. This will greatly improve our understanding of: the extent of the core accretion \cite{Emse21} and gravitational instability \cite{Forg18} formation pathways; planetary sculpting hypotheses for debris disk structures \cite{Bae18}; and the influence of planetary scattering \cite{Mire23}. These insights into giant planet formation and evolution will clarify the impact giant planets have on their terrestrial neighbors (of immediate value for the formulation of HWO), and the overall diversity of planetary system architectures. Discoveries will serve as new planetary benchmarks with temperatures as low as $\sim$250\,K and masses as low as $\sim$0.1\,$M_\mathrm{Jup}$, and can be readily characterised in future JWST cycles through photometry and spectroscopy (Fig.~2). This is a stark contrast to discoveries from NGRST, which will not be amenable to follow-up characterisation. This survey will enable studies of how sub-Jupiter atmospheres are affected by: gravity (cf. higher mass exoplanets / brown dwarfs), irradiation vs. remnant heat from formation (cf. transiting exoplanets), age (cf. Solar System giants), weather (i.e. variability), and low-temperature clouds (e.g., H$_2$O \cite{Morl14}, NH$_3$). Finally, the established ages of the target groups allow companion masses to be determined from evolutionary models, and luminosity differences due to variable initial entropies to be distinguished \cite{Marl14}. 

\vspace{0.2cm}
\textbf{Urgency}: These observations must be performed early on in the lifetime of JWST to enable long term atmospheric characterisation in future cycles, which may also be delayed by any necessary multi-epoch vetting of candidate detections. 

\textbf{Risk/Feasibility}: The occurrence rate of these sub-Jupiters has no direct empirical constraint and the number of detections is uncertain. Nevertheless, non-detections will crucially inform occurrence rates and formation theories. A single verified sub-Jupiter will provide extreme value for exoplanet atmosphere science. 


\textbf{Timeliness}: These science objectives overlap with the ExEP Science Gap list by informing: constraints on exoplanet atmosphere models (SCI-02), architectures of planetary systems (SCI-04), and the influence of giant planets on $\eta_{\oplus}$ (SCI-05).

\textbf{Cannot be accomplished in the normal GO cycle}: 
A large GO program cannot accomplish this survey, and, because the scientific return will happen over many subsequent GO cycles, no GO program can justify the time commitment.
}

\pagebreak
\begin{figure}[!htb]
   \centering
   \includegraphics[width=\linewidth]{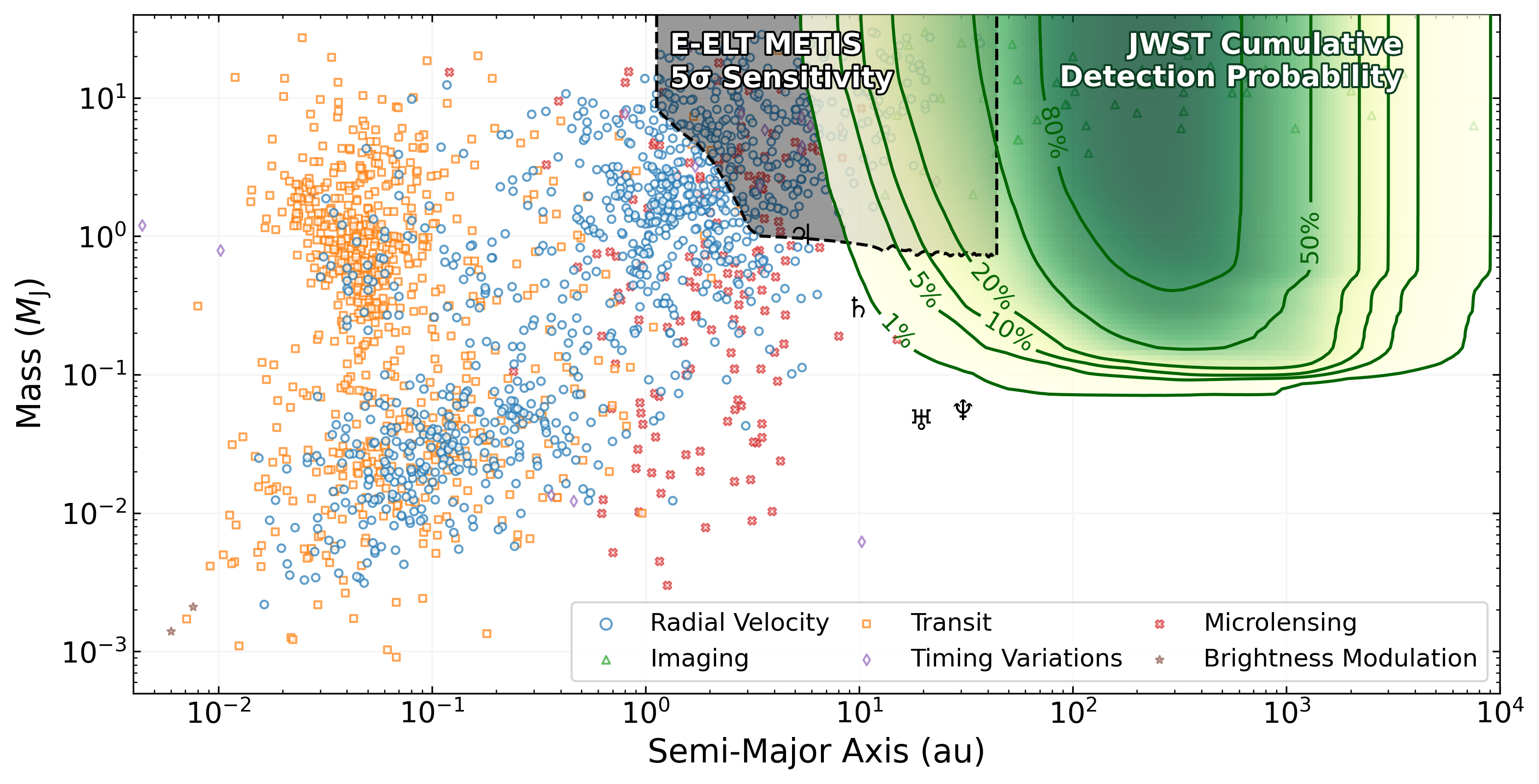} 
    \vspace{-9mm}
    \caption{The known exoplanet population, Solar System planets are marked using their respective symbols. This survey provides unique access to the atmospheres of the wide-separation sub-Jupiter population, as well probing their formation and evolution through the time domain. Existing observatories struggle to push below $\sim$a few ${M_\mathrm{Jup}}$. The E-ELT METIS sensitivity assumes a 10~Myr target at 50~pc. Wide-separation planets detected by NGRST will not be amenable to follow-up characterisation. Example survey is constructed from targets in Taurus Auriga ($\sim$2 Myr), Chamaeleon ($\sim$5 Myr), TW Hya ($\sim$10 Myr), $\beta$ Pictoris ($\sim$26 Myr), and Oceanus ($\sim$500 Myr), others could be considered. The top 25 targets \textit{do not} provide all of the sensitivity below the 10\% contour (for example) - a large survey is required to obtain multiple sub-Jupiter detections.} \vspace{-6mm}
\end{figure}
\vspace{0mm}
\begin{figure}[!htb]
   \centering
   \includegraphics[width=\linewidth]{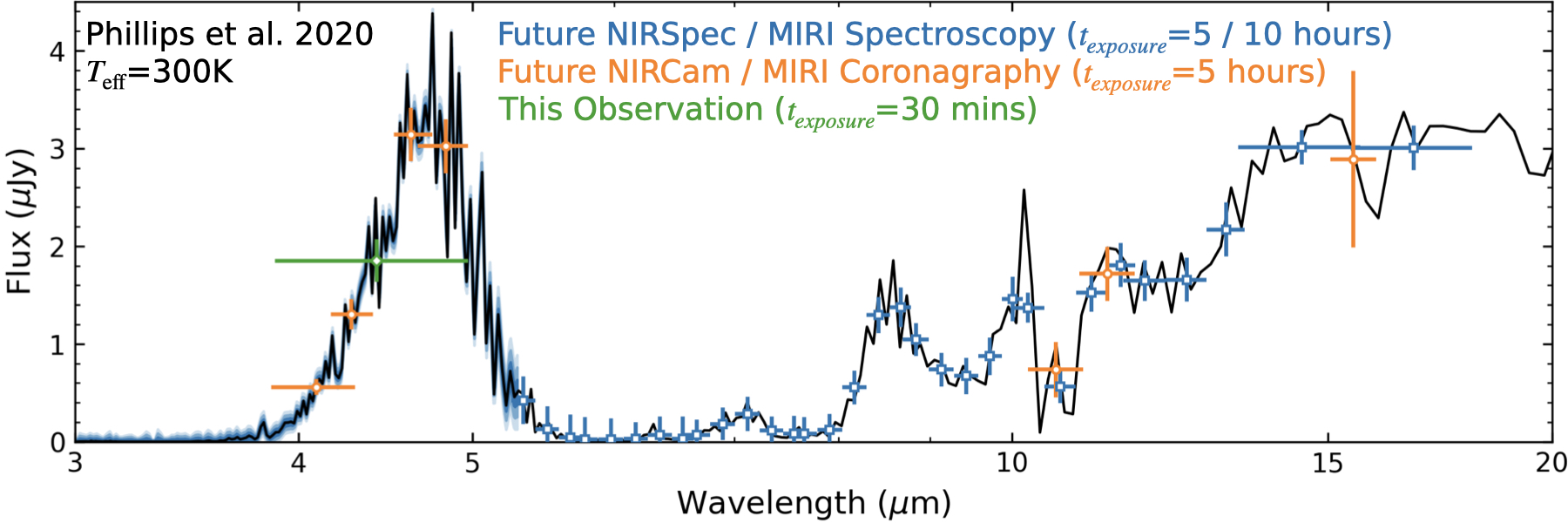}      
   \vspace*{-9mm}
    \caption{Example ETC simulations of future observations of a hypothetical 300~K, Saturn mass, exoplanet. Coronagraphic sensitivity is calculated at 1$''$. Spectroscopic sensitivity is calculated at 3$''$, NIRSpec spectroscopic 1/2/3$\sigma$ regions are shaded, MIRI spectroscopy has been binned to improve signal-to-noise. JWST direct spectroscopy has already been performed on an exoplanet companion at 1.6$''$ separation (GO-2044), and is being pushed to sub-arcsecond separations (GO-3399). Exposure time for coronagraphic observations is per filter.}
\end{figure}

\pagebreak

\printbibliography

\end{document}